# Dampen the Stop-and-Go Traffic with Connected and Automated Vehicles – A Deep Reinforcement Learning Approach*

Liming Jiang[1], Yuanchang Xie[1], Danjue Chen[1], Tienan Li[1], Nicholas G. Evans[2]

*Abstract*— Stop-and-go traffic poses significant challenges to the efficiency and safety of traffic operations, and its impacts and working mechanism have attracted much attention. Recent studies have shown that Connected and Automated Vehicles (CAVs) with carefully designed longitudinal control have the potential to dampen the stop-and-go wave based on simulated vehicle trajectories. This study adopts Deep Reinforcement Learning (DRL) to control the longitudinal behavior of CAVs and utilizes real-world vehicle trajectory data to train the DRL controller. It considers a Human-Driven (HD) vehicle tailed by a CAV, which are then followed by a platoon of HD vehicles. Such an experimental design is to see how the CAV can help to dampen the stop-and-go wave generated by the lead HD vehicle and contribute to smoothing the following HD vehicles' speed profiles. The DRL control is trained using real-world vehicle trajectories, and eventually evaluated using SUMO simulation. The results show that the DRL control decreases the speed oscillation of the CAV by 54% and 8%-28% for those following HD vehicles. Significant fuel consumption savings are also observed. Additionally, the results suggest that CAVs may act as a traffic stabilizer if they choose to behave slightly altruistically.

## I. INTRODUCTION

The stop-and-go traffic shockwaves is an interesting and important phenomenon [1]. Small perturbations in a lead vehicle's speed profile could be amplified as they are passed on to following vehicles and this creates stop-and-go waves broadcast backwards (i.e., traveling upstream) along the road. This phenomenon frequently occurs in high-density traffic conditions because the high density leaves vehicles little space to absorb the wave and traffic flow dynamics tend to be increasingly unstable as density grows. The unnecessary stop-and-go traffic often results in wasted fuel consumption, additional traffic emissions [2], increased likelihood of rear-end crashes [3], and congestion [4].

Automated Vehicles (AVs) technologies have been advancing rapidly in the past 10 years. Accordingly, the research community has shifted its focus from modeling human drivers' behavior to optimize the behavior of AVs. On the other hand, Vehicle-to-Vehicle (V2V) communication technologies allow vehicles to share information with each other in real time and this empowers AVs even more and transforms them into Connected and Automated Vehicles (CAVs). It is concluded [2] that shorter reaction time and better sharing of vehicle maneuver information are among the keys to address the stop-and-go traffic issue. Therefore, CAV appears to be an ideal candidate solution and has attracted much attention recently. However, there are still several open problems for CAV control that are worth exploring, such as human driver behavior cloning [5], [6], CAV assisted merging[7], [8], emission reduction via CAV platooning [9], [10], and developing intelligent controllers that can minimize the propagation of stop-and-go shockwaves in traffic [11]–[13]. In this study, we focus on the last challenge: minimizing the traffic oscillation in a platoon of vehicles.

Previous studies on this subject mostly [11]–[13] use formula-based approaches to control the behavior of CAV with the goal to dampen the stop-and-go traffic. This research adopts a Deep Reinforcement Learning (DRL) approach to see if CAVs can learn optimal control strategies through interacting with human drivers. The highlights of this study are summarized as follows:

1. Instead of using a closed-loop ring network and assuming the location and maneuver information of all vehicles is known as in previous reinforcement learning studies [11], [14], we consider a long straight road segment and take only the CAV and its lead vehicle's state as the algorithm input.

2. Reinforcement learning control models are often trained using simulated data and its effectiveness in practice is sometimes challenged. To test if our DRL model can work in real word, the speed profile of the lead HD vehicle is sampled from field collected vehicle trajectories in naturalistic driving settings.

3. The proposed DRL control only takes a few input parameters based on the state of the CAV and its lead vehicle. Its goal is to dampen the stop-and-go traffic wave, instead of learning optimal Cooperative Adaptive Cruise Control (CACC) strategies as in some previous studies [15], [16] using reinforcement learning.

## II. BACKGROUND

*A. Related work*

Instead of using traditional control theory to come up with formula-based analytic solutions to optimize the behavior of AVs with various objectives, some studies [5], [6], [8], [9], [15], [16] have tried to adopt machine learning algorithms, especially reinforcement learning. The goal of Reinforcement Learning (RL) is to train an optimal longitudinal control policy for CAV through trials and errors. The advantage of using RL is that vehicles can figure out effective control policies implicitly without human directly providing some rule-based

*Research supported by U.S. National Science Foundation (Award #1734521 and #1826162).

[1] Department of Civil and Environmental Engineering, University of Massachusetts (UMass) Lowell, Lowell, MA 01854 USA

[2] Department of Philosophy, UMass Lowell, Lowell, MA 01854 USA
Corresponding Author: Yuanchang Xie, yuanchang_xie@uml.edu
Emails: liming_jiang@student.uml.edu, danjue_chen@uml.edu, tienan_li@student.uml.edu, nicholas_evans@uml.edu

or analytical vehicle control models. By carefully choosing the state representation and reward function, a RL agent (i.e., CAV) is able to learn how to best regulate its longitudinal behavior based on incentives (rewards) received during interactions with other vehicles (the lead HD vehicle in this research).

The first work using RL to control AV in a connected environment was conducted by Desjardins and Chaib-draa [15] The concluded that RL-based control could be a promising approach to ensure a safe longitudinal following behavior of CAV to its front vehicle. After that, RL has been widely adopted in CAV behavior modeling, such as longitudinal control [16], merging [8], [17], and lane-changing decision making [18].

Another interesting study on RL and CAV behavior modeling was done by Vinitsky and Kreidieh [19]. They used different RL algorithms to control mixed-autonomy traffic for merging, uncontrolled intersection and signalized urban grid network scenarios. Their RL algorithms clearly outperformed human driven vehicles, suggesting that RL has great potential for CAV control.

The most relevant study to this paper was conducted by Wu [14]. She considered a closed-loop ring road network, which was loaded with some Human Driven (HD) vehicles and one CAV controlled by the RL algorithm. The RL controlled CAV was assumed to have a global (or complete) view of the environment (i.e., speeds and positions of all vehicles), and the CAV learned to address the stop-and-go traffic by maximizing its reward function, which was defined as the sum of the speeds of each vehicle in the ring network. Although the concept and results of this study are both very interesting and inspiring, assuming a global view of the environment is restrictive and is unrealistic. Also, the RL controller in her study was trained completely based on simulated data, which may not accurately reflect vehicle maneuvers in practice.

Some other relevant studies focused on RL for Cooperative Adaptive Cruise Control (CACC), which is considered a Level 1 automation. CACC controlled vehicles automatically adjust their accelerations given the location, speed, and maneuver information of the front vehicle shared through real-time V2V communication. Most of these studies adopted RL to train CAVs so that they can stably follow the lead vehicle. However, they did not consider using the RL-controlled CAVs to dampen the effects of stop-and-go shockwave to vehicles following them. In other words, these CAVs behave selfishly without considering other vehicles.

CAVs are equipped with advanced sensors. Presumably they can know the surrounding environment better than human drivers and make more informed decisions than human drivers. They can be trained to behave selfishly or a little altruistically (e.g., dampening the stop-and-go traffic wave). An interesting but not fully understood question is how these different behaviors may affect traffic operations and to what extent, and this is what has motivated this study. To our best knowledge, this study is the first attempt to use Deep Reinforcement Learning (DRL) approach for CAV control that builds altruism into the control objective (e.g., dampening stop-and-go wave) and also considers real-world vehicle trajectories instead of simulated ones for training.

*B. Reinforcement Learning (RL)*

With Reinforcement Learning (RL) control, each CAV is treated as a RL agent. The agent learns optimal control policies or strategies through its interactions with the environment (i.e., surrounding vehicles). Good control strategies are rewarded and bad ones are penalized. Over time the agent learns to adjust its behavior to maximize the long-term reward or return. More specifically, at time step $t$ the agent observes a state $S_t$ from its environment. The agent then picks and applies an action $A_t$ to the environment. Because of the action, the environment including the agent would transition into a new state $S_{t+1}$ by following the environment dynamics $\mathcal{P}: \mathcal{S} \times \mathcal{A} \times \mathcal{S} \to \mathbb{R}_+$, where $\mathcal{P}$ is a transitional probability function. A reward is given to the agent based on the new state $S_{t+1}$. The mechanism to assign a reward to an agent is through reward function $r: \mathcal{S} \to \mathbb{R}$. If the long-term discounted return is defined as $\eta(\pi_\theta) = \sum_{i=0}^{T} \gamma^i r_i$, where $\gamma$ is a discounted factor, then the goal of the agent is to learn an optimal control policy that can maximize discounted reward $\pi^* \coloneqq argmax_\theta \eta(\pi_\theta)$.

There are several types of RL algorithms for computing the optimal policy $\pi^*$. Policy-based algorithms try to directly map state to action by learning a policy function $\pi(a|S, \theta)$ parameterized by $\theta$. Value-based algorithm manage to measure how good it is to be at each state by estimating the expected return, then an optimal policy can be derived. policy-based algorithms have the advantages of (1) better convergence property; and (2) more effective in high-dimensional or continuous spaces. But compared to value-based algorithms, they tend to converge to a local optimum rather than a global one and has high variance when updating $\theta$. Actor-critic algorithms [20] share the advantages of both policy-based and value-based methods, and are adopted in this study.

III. METHODOLOGY

*A. Simulation Environment*

The scenario considered in this study is a 70-kilometer single-lane road segment created using SUMO (Simulation of Urban MObility). During simulation, a 10-vehicle platoon is created as shown in Figure 1. The first vehicle is a Human Driven (HD) vehicle that creates the stop-and-go traffic pattern. It is followed by a CAV and eight other HD vehicles. It is assumed that the lead vehicle shares its information with the CAV via V2V communication in real time. The CAV is controlled by a RL agent with the purpose to dampen the stop-and-go shockwave. All other following HD vehicles' behaviors are governed by a modified version of Krauss model [21].

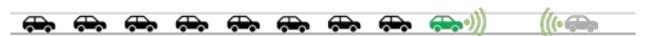

Figure 1. simulation scenario ($1^{st}$ vehicle: stop-and-go pattern; $2^{nd}$ vehicle: CAV controlled by DRL; $3^{rd}$- $10^{th}$ vehicle: human drivers)

Our goal is to ensure that the trained RL agent can tackle realistic car-following tasks. Therefore, the speed profile of the lead vehicle is sampled from the trajectories of a field observed *highD* dataset and reflects real-world driving behavior in congested traffic conditions. The *highD* (The Highway Drone Dataset) [22] contains naturalistic vehicle trajectories recorded on German highways by drones. With advanced computer vision algorithms and high-resolution cameras, *highD* data produces much more accurate vehicle trajectories than other similar well-known datasets such as NGSIM. The *highD* data has a position error of less than 10 centimeters (4 inches) and a high frame frequency of 25 Hz.

All *highD* trajectories were captured on freeways and most of the vehicles were traveling in free-flow mode. Since free-flow traffic does not tell us much about CAV's capability of absorbing the speed oscillation of the lead vehicle, a congested road segment (see Figure 2) in AM peak is hand-picked and only trajectories with a maximum speed less than $18\ m/s$ and a standard deviation of speed more than $2\ m/s$ are selected. These trajectories are concatenated to generate the speed profile of the lead vehicle in this study.

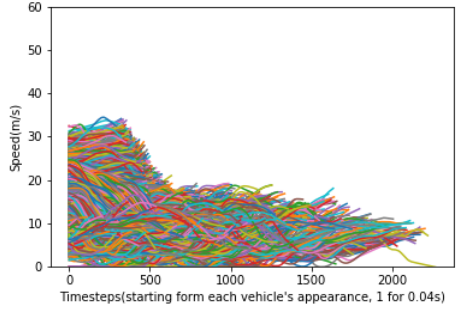

Figure 2. Speed dynamics of vehicles in the selected congested segment

To concatenate the sampled *highD* trajectories, a gentle acceleration rate ($\pm 0.2 m/s^2$) is adopted to stitch these trajectories together. For instance, if the end speed of the previous trajectory is smaller than the start speed of the next trajectory, a positive acceleration rate of $0.2 m/s^2$ is adopted to fill the speed gap of the two-consecutive trajectories. Otherwise, a deceleration rate of $-0.2 m/s^2$ would be adopted.

*B. Parameter Modeling*

In RL, three critical factors are: system state representation, action space definition, and reward function. In this section, the three parameters are described in details.

- State Representation

The system state represents what an agent can actually senses regarding the surrounding environment. It may not be desirable to include all available information. A good state representation design should capture only the necessary information about the environment without including much redundancy. Excluding redundant and trivial information will help to save computation time and improve system reliability.

In this study, we consider the maneuver information of the lead vehicle and the ego vehicle (the RL-controller CAV). We assume they are in a connected environment and there are on-board devices that let these two vehicles share movement dynamics information in real time. Specifically, the state representation is defined as:

$$\{\Delta s, v_{lead}, a_{lead}, v_{ego}, a_{ego}\}$$

where, $\Delta s$ is the distance between the lead vehicle and the ego vehicle. $v_{lead}$ and $a_{lead}$ are the speed and acceleration of the lead vehicle, respectively. Accordingly, $v_{ego}$ and $a_{ego}$ are the speed of acceleration of the ego vehicle, respectively.

- Action Space

Action space defines the range of actions the RL agent can execute at each time step. To reflect the realistic characteristics of regular vehicles. we restrict the acceleration rate (i.e., actions pace) to be in the range of (-3, 2) in $m/s^2$.

- Reward Function

Reward function is a mapping from state representation to a reward received by agent at each time step. Reward function serves as motivations to agent and is the key design feature one can control to regulate agent's behavior. A good design of reward function helps agent learn the intended behavior and facilitates the learning process to converge to an optimum relatively fast.

Our reward function design consists of four main goals. The first goal is for ensuring safety. Depending on the time headway between the lead and ego vehicles, the agent will be given the safety reward defined in Eq. (1).

$$Reward_{headway} = \begin{cases} -100, if\ headway \leq 0 \\ -100 + \sqrt{100^2(1-(x-1)^2)}, \\ \quad if\ 0 < headway \leq 1 \\ 0, if\ headway > 1 \end{cases} \quad (1)$$

Clearly one can design a safe CAV control to dampen the stop-and-go waves by letting the ego vehicle (i.e., the CAV) travel at a constant but very low speed. As long as the ego vehicle's speed is much less than the lead vehicle's average speed, the ego vehicle most likely will not need to decelerate and can maintain a safe headway with the lead vehicle. However, this approach would constantly increase the ego vehicle's gap to the lead vehicle, thus making it a moving bottleneck and increasing the anxiety of drivers behind it. Therefore, another goal is considered to ensure that the ego vehicle maintains a safe headway but also a reasonably fast speed.

In our design, we break down the speed goal into two parts as in Eq. , in which $Reward_{speed}$ is the reward regarding the speed of the ego vehicle, $v_{Ept}$ is the expected speed or speed limit, and $v_{ego}$ is the speed of the ego vehicle. The first part of Eq. (2) encourages the ego vehicle to maintain a high speed, while the second part penalizes the ego vehicle for going slower than the expected speed but does not reward it for going faster than it. By combining the two parts, the goal is for the ego vehicle to catch up with the lead vehicle but not to travel too faster than it.

$$Reward_{speed} = v_{Ept} - Max(0, v_{Ept} - v_{ego}) \quad (2)$$

The third goal is to dampen the speed variation. In other words, when the lead vehicle executes a hard deceleration, the ego vehicle should be able to predict that and take proactive actions (e.g., maintain a large time headway in anticipation of the hard deceleration) so that a hard deceleration is not needed for the ego vehicle and the following HD vehicles would not need to brake hard either.

To achieve this goal, a speed penalty term defined in Eq. (3) is considered if the headway $h$ is smaller than a critical headway value $h_c$. The rationale behind this is that the ego vehicle is not supposed to travel faster than its lead vehicle when it is approaching the lead vehicle. If it does (e.g., in the situation that the lead vehicle decelerates), the ego vehicle should be rewarded by reducing its speed relative to the lead vehicle.

$$Reward_{speeddiff} = (v_{ego} - v_{lead}) * (h - h_c) \quad (3)$$

Where $v_{ego} - v_{lead}$ is speed difference between the ego vehicle and the lead vehicle, $h$ is the current time headway of the ego vehicle, and $h_c$ is the critical headway set to be 1 s in this study. This reward is calculated only if the speed of ego vehicle $v_{ego}$ is less than lead vehicle's speed $v_{lead}$ and headway of ego vehicle $h$ is less than the critical headway $h_c$. Based on this equation, more penalty is given to the DRL agent when it drives faster than its lead vehicle and keep a shorter than critical gap to its lead vehicle.

Another reward function term for achieving the third goal is to penalize large acceleration rates. Large acceleration rates (either negative or positive) should be penalized to ensure smooth driving and help to reduce speed oscillation. This term is defined as follows:

$$Reward_{acc} = -a_{ego}^2 \quad (4)$$

As in Eq. (4), to further penalize large accelerations the acceleration of ego vehicle is squared. The following Eq. (5) is the complete reward function that includes all previously discussed reward terms.

$$\begin{aligned} Reward \\ = \alpha * Reward_{headway} + \beta * Reward_{speed} + \gamma \\ * Reward_{speeddiff} + \delta * Reward_{acc} \end{aligned} \quad (5)$$

In Eq. (5) $\alpha, \beta, \gamma$ and $\delta$ are hyperparameters of this reward function, which specify the weights for each reward terms. After careful hyperparameter tuning, we decide to choose the following values: $\alpha = 1, \beta = 1, \gamma = 1$, and $\delta = 4$

*C. Deep Deterministic Policy Gradient (DDPG)*

Deep learning helps reinforcement learning perform better in complex environments with high-dimensional state and/or action spaces. A representative model of Deep Reinforcement Learning (DRL), Deep Q Network (DQN) [23], is able to solve problems with high-dimensional state spaces. However, it can only handle low-dimensional and discrete action spaces. In our study, the action space for CAV is continuous for a realistic driving agent.

To address DQN's inability to optimize the policy of RL agents in environments with continuous action spaces, Deep Deterministic Policy Gradient (DDPG) was proposed by Google [24] in 2015. DDPG is an actor-critic and model-free algorithm. A brief introduction of DDPG is provided below, and one may refer to the original paper for notations and other detailed information.

In general, value-based RL methods often suffer from poor convergence, while policy-based RL methods tend to converge to local maximus and suffer from high variance and sample inefficiency. Actor-critic RL methods combine the advantages offered by both value-based and policy-based methods by employing an actor (to execute an action) and a critic (to evaluate the action from the actor). This allows actor-critic RL to be more sample efficient [24].

After a minibatch of $N$ transitions $(s_i, a_i, r_i, s_{i+1})$ is sampled from the replay buffer. The target $y_i$ is calculated as

$$y_i = r_i + \gamma Q^i(s_{i+1}, \mu'\left(s_{i+1}\middle|\theta^{\mu'}\right)\theta^{Q'}) \quad (1)$$

Then the model of critic is updated by minimizing the loss:

$$L = \frac{1}{N}\sum_i\left(y_i - Q(s_i, a_i|\theta^Q)\right)^2 \quad (2)$$

Accordingly, the sampled policy gradient is adopted to update the actor policy.

$$\begin{aligned} \nabla_{\theta^\mu} J \\ \approx \frac{1}{N}\sum_i \nabla_a Q(s, a|\theta^Q)|_{s=s_i, a=\mu(s_i)} \nabla_{\theta^\mu}\mu(s|\theta^\mu)|_{s_i} \end{aligned} \quad (3)$$

IV. RESULTS

In this section, key results are presented and compared to the baseline model both qualitatively and quantitatively. Additionally, a behavior analysis is conducted to understand the vehicle behavior generated by the proposed model.

*A. Evaluation Methodology*

The trained CAV DRL control is coded and evaluated in SUMO simulation. Another identical set of SUMO simulation is also conducted except that only Human Drivers (HD) are considered. In other words, we replace the DRL controlled ego vehicle with a HD vehicle. The HD scenario serves as the baseline to demonstrate the superiority of the proposed DRL model.

The following rolling mean and standard deviation of speed are adopted to measure the capability of DRL controlled CAVs to absorb hard decelerations of the lead vehicle.

$$\bar{x}_k^T = \frac{1}{T}\sum_{i=k}^{k+T} x_i \quad (4)$$

$$s_k^T = \sqrt{\frac{1}{T-1}\sum_{i=k}^{k+T}(x_i - \bar{x}_k^T)^2} \quad (5)$$

Where $T$ is the rolling time window length, $\bar{x}_k^T$ is the average speed starting at the $k^{th}$ time step of length $T$, and $s_k^T$ is the rolling standard deviation of speeds starting at the $k^{th}$ time step. To measure the overall speed variation over the entire evaluation period, the average of all rolling standard deviations $s_i^T$ is utilized as defined below.

$$\bar{s}^T = \frac{1}{end - start - T - 1} \sum_{i=start}^{end-T} s_i^T \qquad (6)$$

### B. Experiment Results

The total evaluation simulation run time is about two hours. Simulated vehicle trajectories from a randomly selected 2.5-minute period are extracted (See Figure 3) to compare the performances of the proposed model and the baseline. Based on the trajectories, vehicle speed and acceleration profiles are also plotted and presented in Figures 4 and 5.

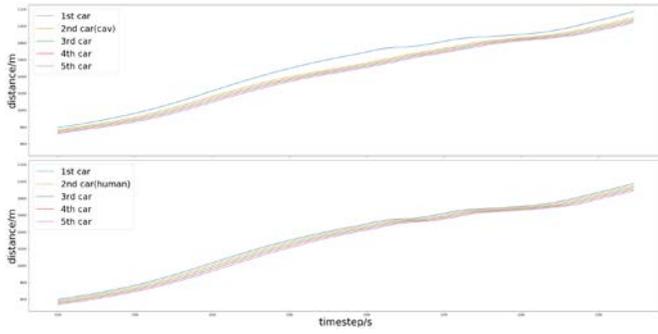

Figure 3. Trajectories (Top: DRL agent as 2nd car; Bottom: human driver as 2nd car)

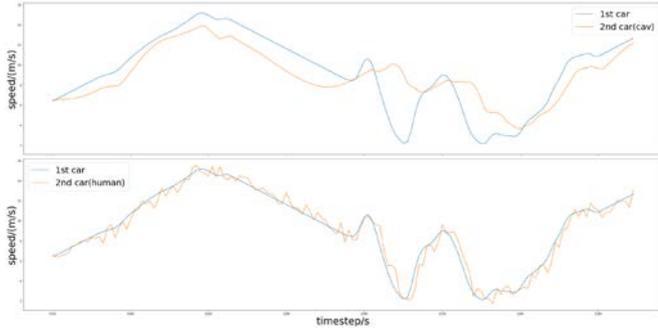

Figure 4. Speed profiles of the first two cars (Top: DRL agent as 2nd car; Bottom: human driver as 2nd car)

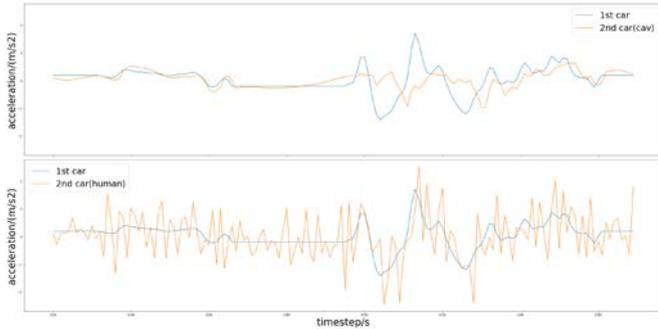

Figure 5. Acceleration profiles of the first two cars (Top: DRL agent as the 2$^{nd}$ car; Bottom: human driver as the 2$^{nd}$ car)

As can be seen in the trajectory figure (Figure 3), the DRL-controlled CAV tends to keep a longer distance (about 50 meters on average with a standard deviation of 30 meters) to its front car than human drivers. Although this distance is not explicitly defined by the reward function in our design, the DRL agent figures out by itself that in order to avoid hard and/or frequent accelerations/decelerations, it needs to increase the gap to its lead vehicle and use that gap as a buffer zone to absorb the stop-and-go shockwave.

Another interesting finding can be observed from the speed profile diagram (Figure 4). For a HD follower of the lead vehicle, the HD vehicle tends to copy the behavior of its lead vehicle and further exaggerates the speed oscillation. For example (in Figure 4), every time the lead vehicle decelerates to a speed $v$, the following HD vehicle tends to decelerate to an even smaller speed compared to $v$. While the CAV tends to dampen the speed oscillation of the lead vehicle and take a less extreme action compared to its lead vehicle. The acceleration profiles for CAV and HD vehicle (Figure 5) also show that CAV is able to keep the acceleration rates within a much smaller range.

After the above comparison, a valid question is that is the DRL control going to increase the overall travel time. For example, the CAV travels at a very low and constant speed. Although this can lead to a very smooth trajectory, it will take the CAV much long time to travel the same distance than a HD vehicle with a stop-and-go trajectory. Based on the simulated results, the DRL-controlled CAV is able to not only absorb the stop-and-go shockwave created by the lead vehicle, but also travel at the same average journey speed as a HD vehicle. Note that both the CAV and HD vehicle speeds are constrained by the lead vehicle.

To quantify the effects of how stop-and-go waves get dampened with CAV and HD vehicle, the average rolling speed standard deviations for vehicles at different positions are calculated and presented in Table 1.

TABLE I. $\bar{s}^T$ FOR DIFFERENT VEHICLES GIVEN CAV AND HD

| Vehicle Order | 1st | 2nd | 3rd | 4th | 5th |
|---|---|---|---|---|---|
| $\bar{s}^T$ (DRL agent) | 0.34 | 0.27 (-54%) | 0.55 (-28%) | 0.70 (-16%) | 0.78 (-14%) |
| $\bar{s}^T$ (all human) | 0.34 | 0.58 | 0.76 | 0.83 | 0.91 |
| Vehicle Order | 6th | 7th | 8th | 9th | 10th |
| $\bar{s}^T$ (DRL agent) | 0.84 (-11%) | 0.92 (-8%) | 0.92 (-13%) | 0.93 (-17%) | 0.95 (-14%) |
| $\bar{s}^T$ (all human) | 0.95 | 1.00 | 1.06 | 1.13 | 1.11 |

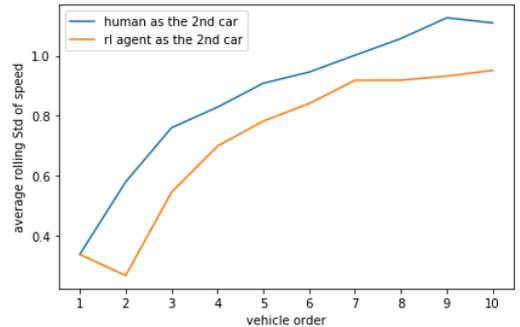

Figure 6. Speed oscillations of different vehicles given CAV and HD

In Table 1, the first vehicles in both CAV and HD scenarios have the same speed oscillation as their trajectories are sampled from the *highD* dataset and are identical. For the 2nd vehicle in the platoon, the CAV is able to absorb the speed oscillation by 54% compared to its HD vehicle counterpart. The effect on the 3rd vehicle in the CAV scenario is less significant because the 3rd vehicle is controlled by a human driver. Nevertheless, it still has a 28% reduction in speed oscillation compared to the HD scenario. For the remaining vehicles in the platoon, the speed oscillation reduction benefits are in the range of 8%-17%. In sum, the oscillation reduction benefits reach the peak (over 50% reduction) for the 2nd vehicle in the platoon, and drop to the somewhere near 8% as the stop-and-go wave propagates to the 7th vehicle, and increase slightly afterwards.

The fuel consumption reduction benefits are also studied. The emission model HBEFA [25] is used to quantify the fuel consumption for each vehicle in both scenarios, and the comparison results are plotted in Figure 7. The overall patterns for fuel consumption saving and speed oscillation reduction are similar.

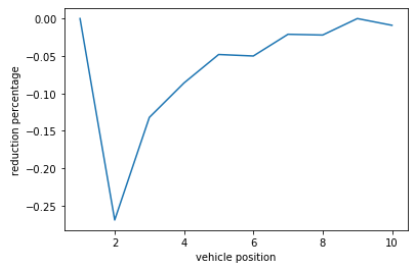

Figure 7. Fuel consumption of CAV scenario compared to HD scenario.

## V. CONCLUSION

This study shows that DRL is a promising approach to dampen stop-and-go traffic and generates significant safety and environmental benefits in terms of speed variation and fuel consumption reductions, respectively. These benefits are not only for the ego CAV vehicle, but also for other human driven vehicles following it. This brings up an interesting question for future research: should CAVs behave in its own interest only or altruistically? Also, more work can be done considering multiple DRL-controlled CAVs. Additionally, it would be interesting to model the performance of the DRL model in a multi-lane environment.